\documentclass[11pt,A4paper]{article}
\usepackage{graphicx}
\title{Heavy ion collisions in the 1A GeV regime: how well can we join up to
astrophysics?}
\author{
W. Reisdorf for the FOPI collaboration \\
\emph{GSI, Planckstr,, Darmstadt, Germany} \\
}

\date{}  %no-date, default

\begin{document}
\maketitle

\begin{abstract}
 The derivation of information useful for understanding the physics
inside compact stars from HIC observations is a difficult task. 
Complications due to finite size, different chemistry, non-adiabatic
compression, incomplete stopping and structural effects must be overcome.
%This can only be convincingly achieved by a successful simulation of many 
%experimental observables varying incident energy, system size and isospin
%content.
Using now available systematic FOPI data in the SIS energy range we try to
trace the path to take.
%Specific attention is devoted to isospin observables.
\end{abstract}

%-----------------------------------------------------------------------------
\begin{figure}[h]
\centering
\includegraphics[width=0.46\textwidth]{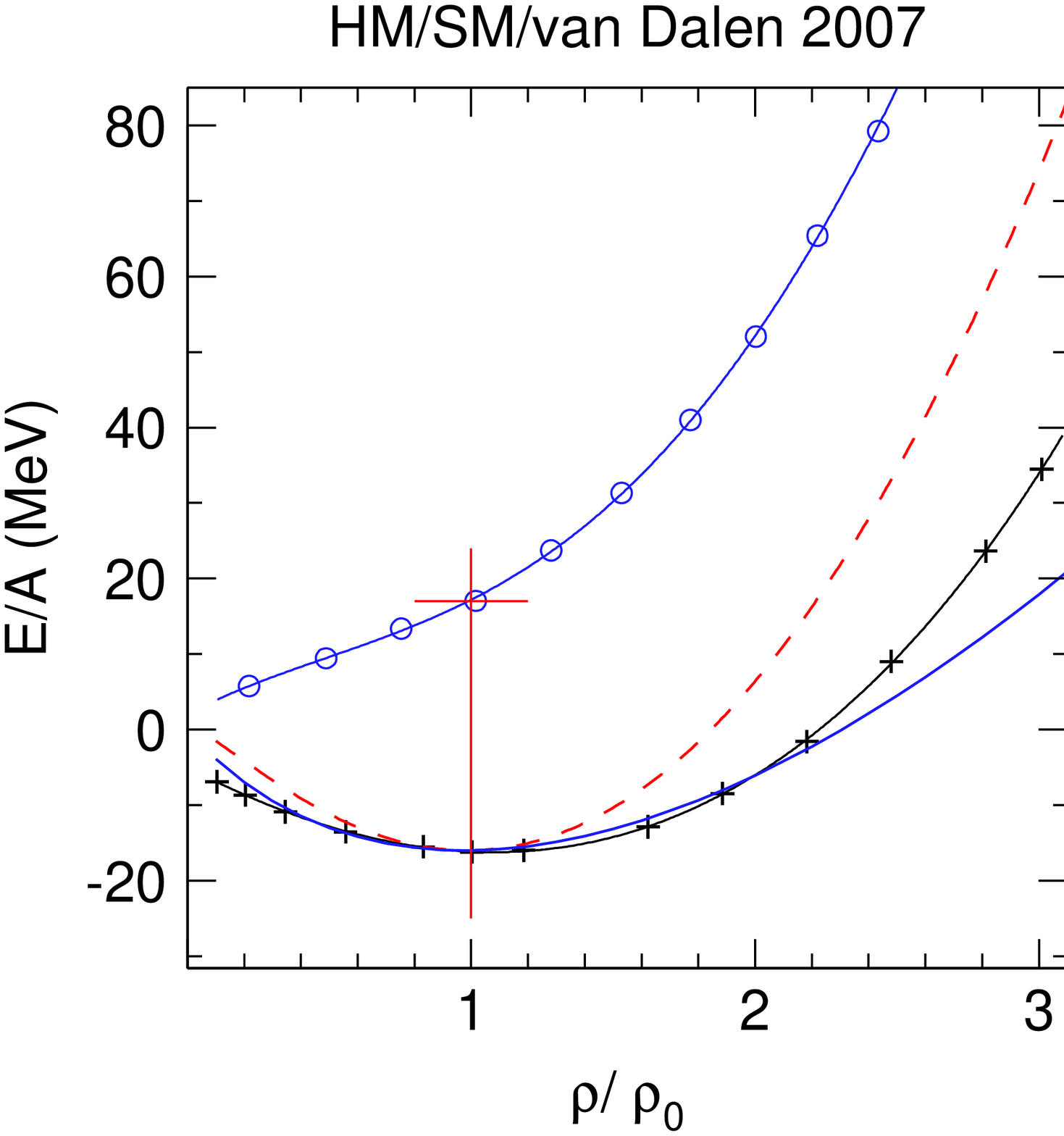}
\includegraphics[width=0.44\textwidth]{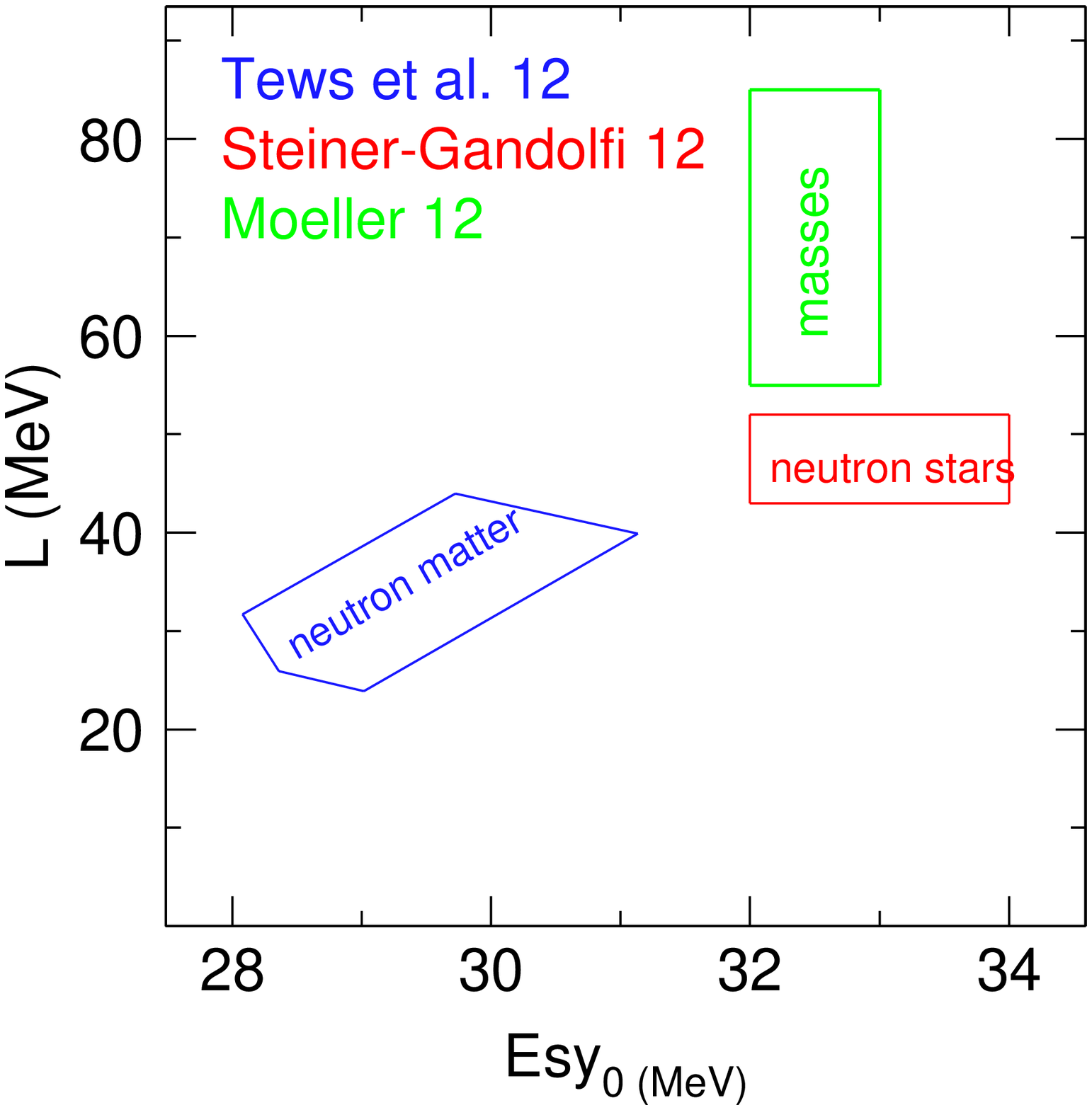}
\caption{Various nuclear matter EoS (left) and constraints (right)}
\label{fig01}
\end{figure}

There now exists an extensive set of data on heavy ion reactions in the
$1A$ GeV range \cite{rei12}. 
In the sequel we confront the data with IQMD \cite{hart98} simulations.
The two options of purely phenomenological {\it cold} nuclear EoS that we use
are plotted in Fig.1 and confronted with a 'microscopic' 
(Dirac-Brueckner-Hartree-Fock, DBHF) calculation \cite{dalen07}
for symmetric matter.
It is seen that in the density range relevant for SIS energies (up to
$\rho/\rho_0=2.5$) our 'soft' version, SM, is rather close to the theoretical
calculation. 
We will see in the sequel that FOPI data are strongly favouring the
SM version (full blue) over the stiff, HM, version (dashed).

Also included from the same theoretical work is the
cold EoS for pure neutron matter.
It is  not possible in the laboratory to determine directly the
neutron matter EoS. We have to rely on theoretical help.
The adequacy of the theory, in turn, can be tested by
confrontation with high quality experimental data constraining
 the symmetric matter EoS.
Recent constraints on the EoS parameters $L$ and $Esy_0$ from
theoretical efforts \cite{tews12}, nuclear masses \cite{moe12}
 and neutron star data
\cite{stei12} reflect incompatibilties associated with different physics
sensitivities, see Fig.1 right.

In Fig.2 we show a sample of proton yield and flow  data from our Collaboration
(black dots with error bars) together with simulations using IQMD with the
stiff version of the EOS (HM, red dashed) and the soft version (SM, blue
full).

\begin{figure}[h]
\centering
\includegraphics[width=0.70\textwidth]{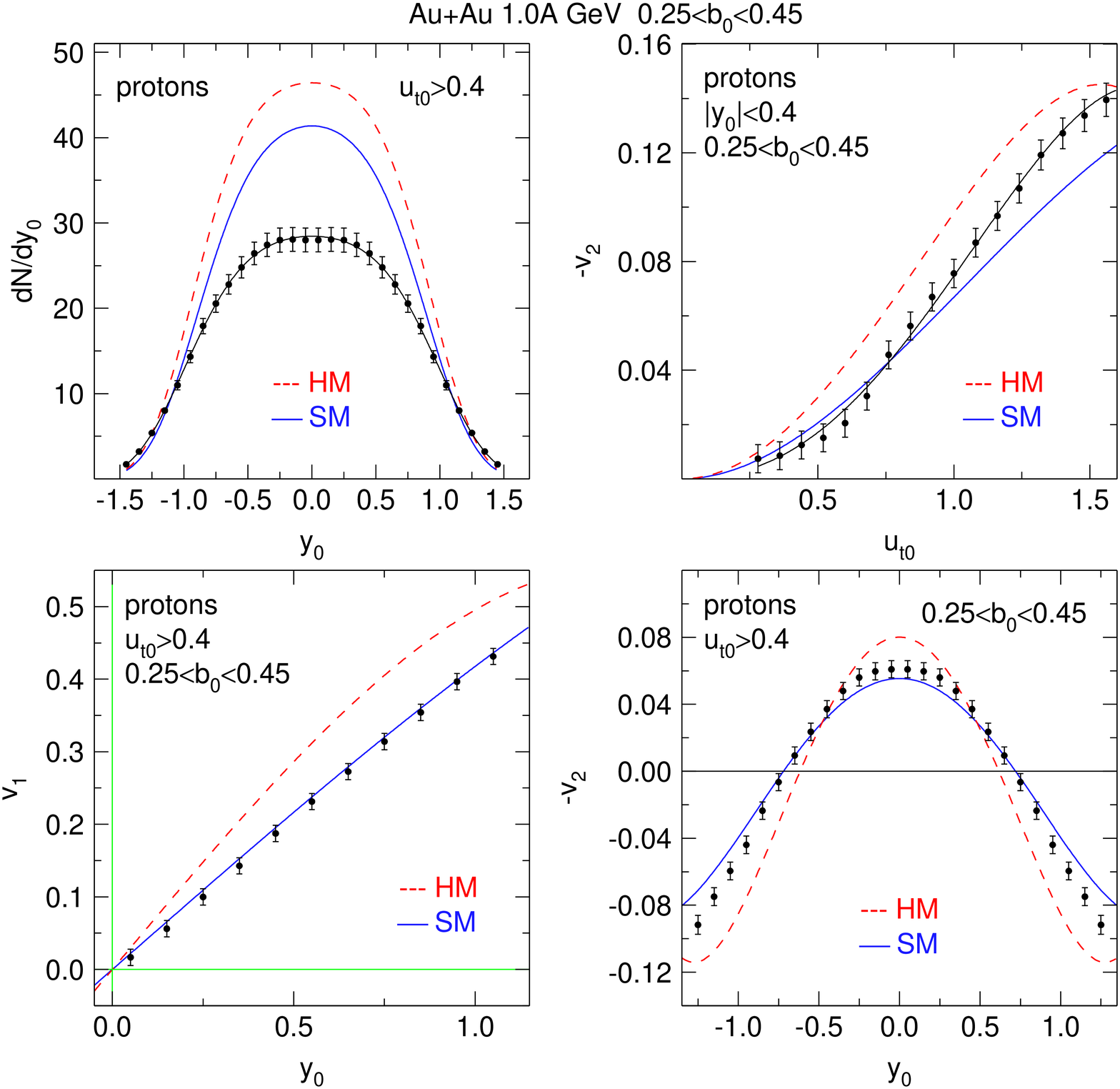}
\caption{Proton rapidity and flow data and IQMD-SM/HM simulations}
\label{fig01}
\end{figure}

It can be seen that the three projections shown are best described by the SM
version: see the rapidity ($y$) dependences of
the directed, $v1$, and the elliptic ($-v_2$) flow in the two lower panels
and the $p_t/m=u_t$ dependence of the elliptic flow in the upper right panel.
As IQMD underestimates clusterization it overpredicts single nucleon (proton)
yields (upper left panel). 
Notice a moderate, but still remarkable dependence on the EoS, however.

Taking a closer look at $-v_2(y_0)$ (we use the index 0 to indicate scaling
with the beam parameters \cite{rei12}) we see that the predicted shape
is sensitive to the EoS {\it in the full rapidity range}.
To take advantage of this feature we introduce a quantity dubbed $v_{2n}$
defined by $v_2n=-v_{20}+|v_{22}|$ where the parameters are fixed by a fit
to the flow data using $v_2(y_0)=v_{20}+v_{22}\cdot y_0^2$ in the scaled rapidity
range $|y_0|<0.8$.   

The result for Au+Au between $0.4A$ and $1.5A$ GeV is shown in Fig.3
for protons (lower left) and deuterons (lower right), tritons (upper left)
and $^3$He.
%----------------------------------------------------------------------

\begin{figure}[h]
\centering
\includegraphics[width=0.85\textwidth]{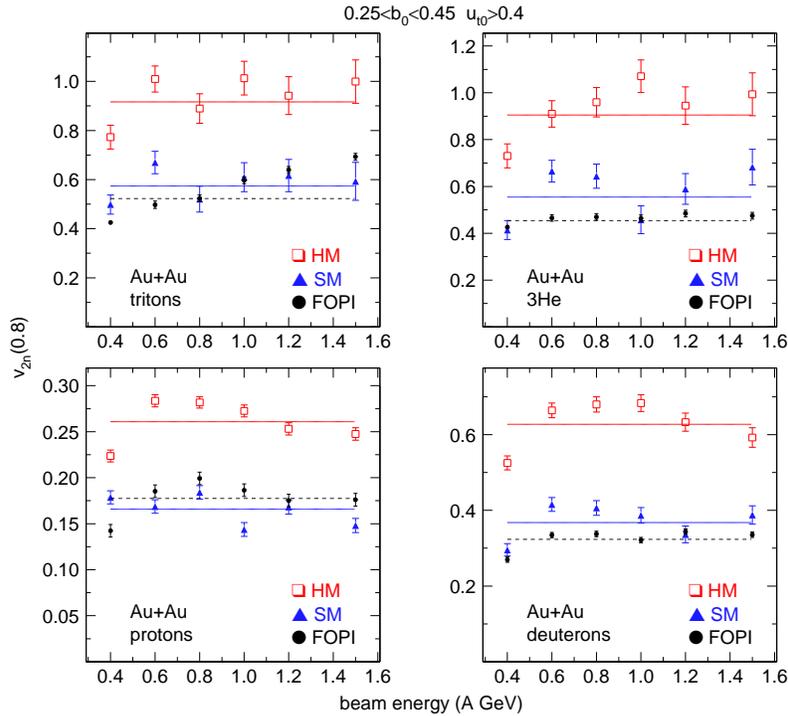}
\caption{Elliptic flow $v_{2n}$ for protons, deuterons, tritons, $^3$He}
\label{fig01}
\end{figure}

As the beam energy dependences are rather weak, we indicate the average
behaviour by straight lines. The comparison of the data for $v_{2n}$ with
the calculations shows a rather convincing preference for SM!
The sensitivity is large: there is a factor 1.6 between HM and SM, a
difference exceeding significantly the indicated experimental error bars.
This strongly supports the T\"{u}bingen calculation (Fig.1).

\begin{figure}[h]
\centering
\includegraphics[width=0.85\textwidth]{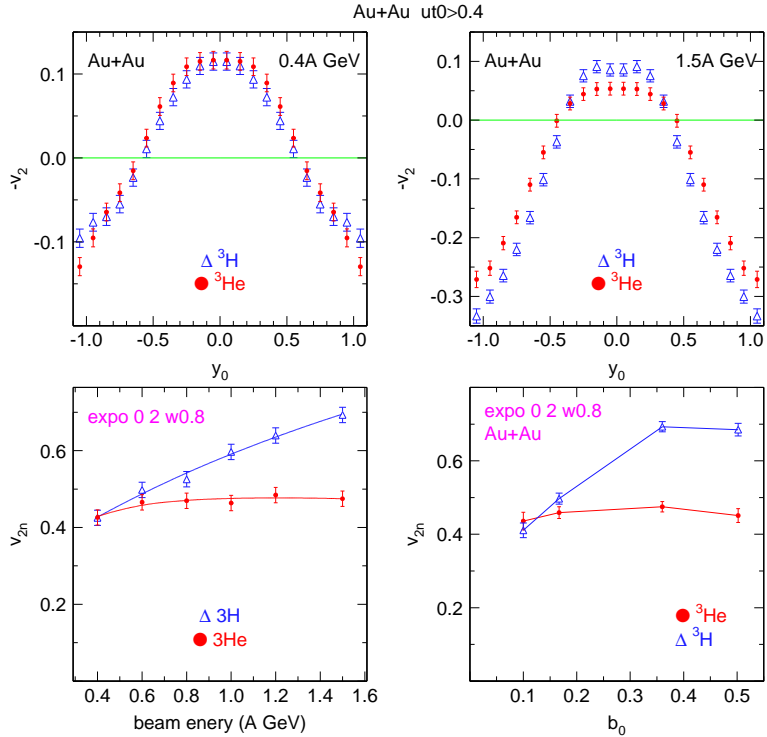}
\caption{Studies of $v_2$ and $v_{2n}$ for the two mass three isotopes}
\label{fig01}
\end{figure}

Including mass three clusters (besides protons and deuterons) leads to the
same conclusions.
In view of high interests in isospin dependences it is worth looking in more
detail at elliptic flow data of $^3$H and $^3$He: see the two upper panels
in Fig.4 for Au+Au at $E/A=0.4A$ (left) and $1.5A$ GeV (right).
While there is no significant (within error bars) difference at the lower
beam energy we see a remarkable effect at $1.5A$ GeV: the shape difference
in $-v_2(y_0)$ is reminiscent of the SM/HM shape difference seen in Fig.2.
We therefore use again $v_{2n}$ to systematize this isotopic difference in
terms of a single parameter: see the two lower panels showing the
energy dependence for both isotopes (left) and,  for the $1.5A$ GeV data,
the centrality dependence in terms of the scaled impact parameter $b_0$ 
 \cite{rei12}.

These observations, so far, are not reproduced by our IQMD version.
Considering the limitation of the isotopic split
 to larger $b_0$ and higher $E/A$,
we suggest unaccounted for momentum dependences and connection to 
$\Delta$ formation in an asymmetric medium.
For future clarification of the latter we expect our pion yield and flow data
\cite{rei12} to be helpful.

Our conclusions concerning preference, in the SIS energy range, of
a 'soft' EoS (see Fig.1) are in line with earlier findings using
the comparison of $K^+$ yield data varying system size or centrality.
A sample of such data \cite{foerster07} is shown in the right panel of Fig.5.
Once cluster formation is better understood, there is some chance,
that conclusions on the EoS can also be derived from system dependences
(size and isospin) of various clusters, see
the various panels in Fig.5.
There is evidence for more efficient cooling (condensation)
 if the achieved density was
higher, i.e. for more massive systems.
For pions the increased production in softer, denser systems, is
compensated out by the final cooling before freezeout. There is a loss
of memory for the high density phase here.

\begin{figure}[t]
\centering
\includegraphics[width=0.58\textwidth]{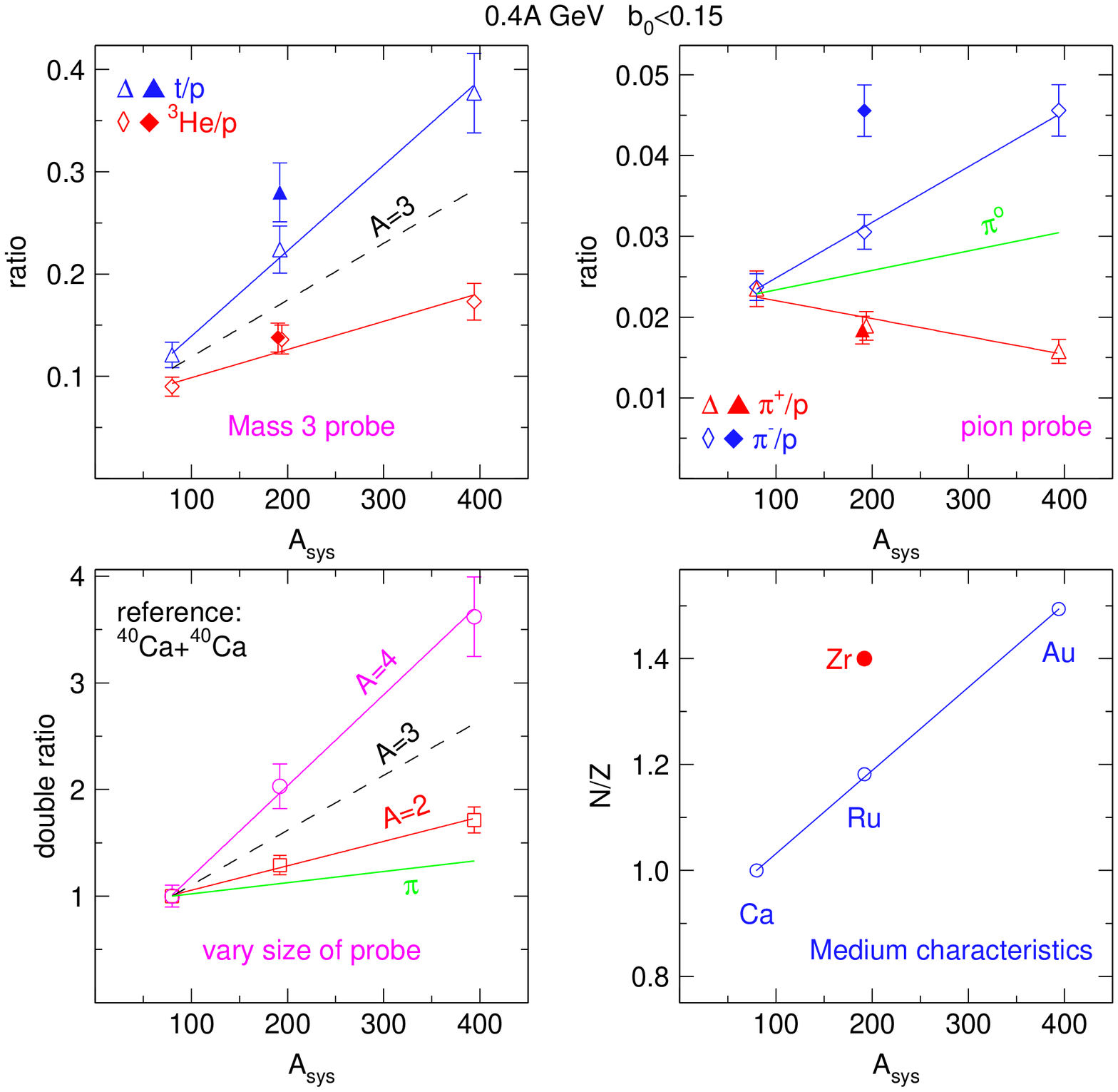}
\includegraphics[width=0.37\textwidth]{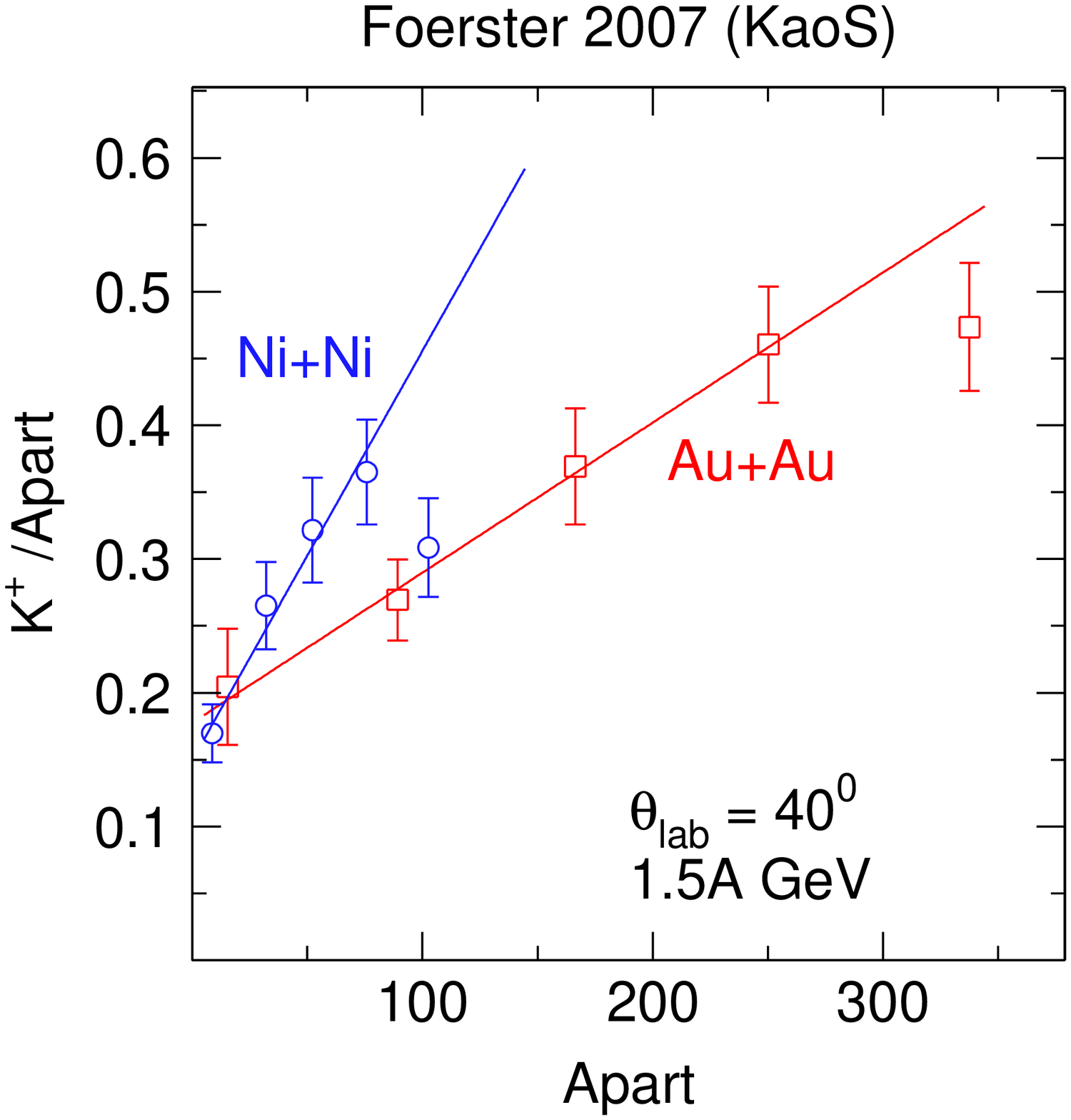}
\caption{System size dependences of various indicated ejectiles}
\label{fig01}
\end{figure}

To conclude, heavy ion data obtained at SIS, represent by now rather
convincing constraints for the EoS of nuclear matter in the density range
up to $\rho=2.5\rho_0$.

\end{document}